\documentclass[12pt]{iopart}
\usepackage{amsmath,graphicx,amssymb}
\begin{document}
\title{Transmittivity measurements by means of squeezed vacuum light}
\author{V. D'Auria$^{\$\dagger}$, C. de Lisio$^{\$\dagger}$, 
A. Porzio$^{\$}$, S. Solimeno$^{\$\dagger}$}
\address{$\$$ Coherentia--CNR--INFM \\ and \\ 
$\dagger $ Dipartimento di Scienze Fisiche, Univ.
\textquotedblleft Federico II\textquotedblright\,\\
Compl. Univ. Monte Sant'Angelo, 80126 Napoli, Italia}
\author{Matteo G.A. Paris}
\address{Dipartimento di Fisica dell'Universit\`{a} di Milano, Italia.} 
\begin{abstract}
A method for measuring the transmittivity of optical samples by using
squeezed--vacuum radiation is illustrated. A squeezed vacuum field generated
by a below--threshold optical parametric oscillator
is propagated through a nondispersive medium and detected by a
homodyne apparatus. The variance of the detected quadrature is used for
measuring the transmittivity. With this method it is drastically reduced the number
of photons passing through the sample during the measurement interval.
The results of some tests are reported.
\end{abstract}

\section{Introduction}

Vacuum fluctuations of electro--magnetic (\textit{e.m.}) fields have been
the ultimate limit on the precision of optical measurements until the advent
of squeezed light. Since then, many attempts have been made for reducing the
shot noise level blurring several types of signals. Caves \cite{caves} first
proposed to combine coherent and squeezed vacuum radiation for overcoming
the quantum limit in gravitational wave antennas. Grangier et al. \cite
{grangier} up-graded a polarization interferometer by injecting a squeezed
vacuum through a dark port. Polzik et al. \cite{polzik} provided stunning
evidence that a gain of some dB over the standard quantum limit is achieved
in the resonant interaction of atoms with squeezed light. Their experiment
was performed by combining in a well defined phase relation a coherent field
with the output of an optical parametric oscillator (OPO) operating
below--threshold.

Other experiments have exploited the correlation between twin beams for
reducing the noise level of the probe field \cite{lane,rarity,shelby,fabre}.
The principles of these measurements were highlighted in Ref. \cite{lane},
where it was recognized that the losses occurring in one beam can be
inferred from those relative to the other one (see also \cite
{shelby,fabre,apb01}).

In this communication a method for probing the transmittivity $T$ of a
sample with squeezed vacuum radiation is discussed. Standard methods rely on
direct measurement of the radiation intensity entering and leaving the
sample. Sufficient accuracy can be achieved by using beams so intense to
contrast the shot--noise, although, in some circumstances, using high input
intensity is either not useful (in case of very low absorption) or unwise
(strongly non-linear materials or samples whose structure may be altered by
intense photon fluxes).

In alternative to the above schemes \cite{caves,grangier,polzik} in the
proposed method the sample is irradiated with a squeezed vacuum field. Then,
the emerging one is combined with a coherent one (local oscillator, LO) in a
balanced homodyne detector measuring the fluctuations in a suitable spectral
range. The interaction of the squeezed vacuum with the sample modifies the
spectrum of the homodyne current by changing its variance. Hence, the
transmittivity is determined by measuring the variance changes. The main
advantage of this method is a very low number of photons interacting with
the sample.

Below--threshold degenerate OPOs produce \textit{e.m.} radiation represented
by a combination of squeezed vacuum and thermal components
(squeezed--thermal--vacuum states, STV) with a Gaussian statistics. The OPO
working conditions determine the STV state properties \cite{Marian,Man'ko}.
The propagation through non resonant media transforms such a state into
another STV one with different variances of the field quadratures $\Delta
X_{\phi }^{2}$. The change of $\Delta X_{\phi }^{2}$ is used for measuring
the transmittivity $T$ (Section 2).

The squeezed radiation is analyzed by a balanced homodyne detector providing
the field quadratures $X_{\phi }=\frac{1}{2}\left( ae^{-i\phi }+a^{\dag
}e^{i\phi }\right) $ via the controlled interference between the STV state
and a strong coherent LO of relative phase $\phi $. Since the detected
signal is proportional to $X_{\phi }$ times the LO amplitude, the detection
is efficient also in case of very weak beams, as in the present case.
Consequently, the effects of the SNR on the accuracy can be disregarded.

Essential to this method is the use of a Gaussian distributed quadrature $%
X_{\phi }$. This means that for testing the method it is necessary to
preliminarily measure the distribution function by sampling $X_{\phi }$ an
adequate number $N$ of times. In alternative, it is also possible to
determine the whole Wigner function with quantum homodyne tomography (QHT) 
\cite{QHT} using samples uniformly distributed over the whole interval $%
\left( 0,2\pi \right)$. Distributing $N$ samples in the interval $\left(
0,2\pi \right)$ reduces the accuracy of only a few percent. This slight loss
is largely compensated by a three dimensional characterization of the STV
state in the phase space.

Aim of this communication is to assess the feasibility of this scheme by
testing the validity of two main assumptions, namely $i$) the generation of
Gaussian STV states by a below--threshold OPO, and $ii$) the description of
the absorption process as a simple scaling of the P-representation of the
STV state. The dependence of the accuracy of the proposed method on the STV
state parameters is also examined. Moreover, the method accuracy is compared
with that achievable with standard techniques. Some measurements carried out
with a below--threshold type--I Lithium Niobate (LNB) OPO at $\lambda =1064$
nm, typically generating few pW STV states are illustrated.

The paper is organized as follows. In the next section the properties of the
STV states undergoing lossy propagation are discussed. Then, in Section 3,
the accuracy of the measurement of $T$ based on this method is compared with
that of different techniques. Section 4 is dedicated to the description of
the experimental tests. Eventually, in Section 5, conclusions are drawn.

\section{Generation and propagation of STV states}

\begin{figure}[h]
\setlength{\unitlength}{1mm}
\par
\begin{center}
\includegraphics[width=0.7\textwidth]{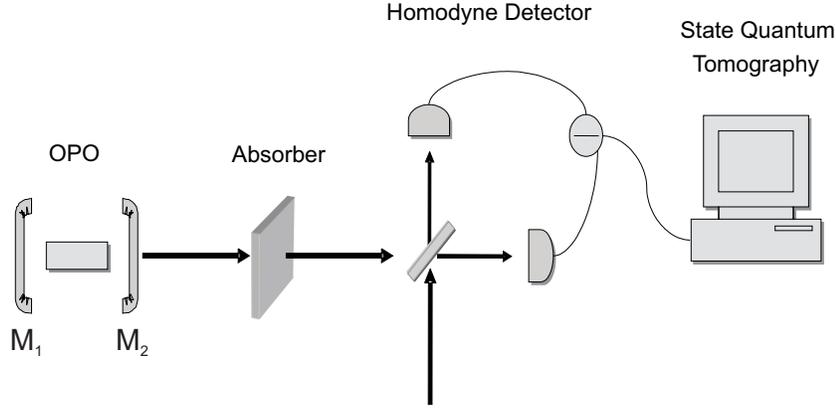}
\end{center}
\par
\vspace{-.5cm}
\caption{Schematic of the OPO cavity and the experimental setup. The STV
states, generated by the OPO, pass through a sample of variable
transmittivity $T$ and then are characterized by a homodyne detector. State
parameters are recovered by QHT data processing.}
\end{figure}

The field generated by a below--threshold OPO satisfies the Langevin
equation:

\begin{eqnarray}
\frac{d}{dt}\left[ 
\begin{array}{c}
a \\ 
a^{\dag }
\end{array}
\right] &=&\left( \gamma +\delta \gamma \left( t\right) \right) \left[ 
\begin{array}{c}
a^{\dagger } \\ 
a
\end{array}
\right] -\left[ 
\begin{array}{c}
\left( \kappa +i\psi +i\delta \psi \left( t\right) \right) a \\ 
\left( \kappa -i\psi -i\delta \psi \left( t\right) \right) a^{\dag }
\end{array}
\right]  \nonumber \\
&&+\sqrt{2\kappa _{1}}\left[ 
\begin{array}{c}
a_{1}^{in} \\ 
a_{1}^{in\dag }
\end{array}
\right] +\sqrt{2\kappa _{2}}\left[ 
\begin{array}{c}
a_{2}^{in} \\ 
a_{2}^{in\dag }
\end{array}
\right] \,,  \label{Langevin OPO}
\end{eqnarray}
where $a_{1}^{in}$ is the noise entering the cavity through the output
mirror $M_{1}$ (see Fig. 1) and $a_{2}^{in}$ represents both the noise
entering through the input mirror $M_{2}$ and the crystal contribution. The
quantities $\kappa _{1}$, $\kappa _{2}$ are damping coefficients, whereas $%
\kappa =\kappa_{1}+\kappa_{2}$. The parametric gain is the sum of a
stationary mean value, $\gamma$, and a small fluctuating contribution, $%
\delta \gamma\left( t\right)$. Similarly, $\psi$ and $\delta \psi \left(
t\right)$ are the mean cavity detuning and its fluctuations, respectively.
In this context, the ratio $\mathcal{E}=\gamma^{2}/\kappa^{2}$ represents
the distance of the actual operating condition from the OPO threshold power,
while $\kappa_{1}/\kappa$ is the so called coupling efficiency.

In the frequency domain the variance $\Delta X^{2}$ of the output quadrature 
$X=X_{\phi =0}$ reads as: 
\begin{eqnarray}
\fl
\Delta X^{2}=\frac{\left| \kappa ^{2}-\gamma ^{2}-\left( \omega ^{2}-\psi
^{2}\right) +i2\omega \kappa -2\kappa _{1}\left( \kappa +i\left( \omega
-\psi \right) +\gamma \right) \right| ^{2}+4\kappa _{1}\kappa _{2}\left|
\kappa +i\left( \omega -\psi \right) +\gamma \right| ^{2}}{4\left| \kappa
^{2}-\gamma ^{2}-\left( \omega ^{2}-\psi ^{2}\right) +i2\omega \kappa
\right| ^{2}}\,, 
\end{eqnarray}
where $\omega $ is the frequency offset from the optical frequency $\omega
_{0}$, and $\delta \gamma \left( t\right) $ and $\delta \psi \left( t\right) 
$ have been neglected. The variance $\Delta Y^{2}$ ($Y=X_{\phi =\pi /2}$) is
given by a similar expression with $\gamma $ replaced by $-\gamma $.

For a single--input cavity ($\kappa _{2}=0$), and for $\psi =0$, the product 
$16\Delta X^{2}\Delta Y^{2}$ reduces to unity, corresponding to a minimum
uncertainty state. In general, this condition is no more satisfied for
double--ended cavities or non--zero detuning or lossy crystals. Then, it is
worth characterizing the OPO output at the sampled frequency $\omega $ by
means of the adimensional parameters: 
\begin{eqnarray}
n_{th} &=&2\left( \sqrt{\Delta X^{2}\Delta Y^{2}}-\frac{1}{4}\right) 
\nonumber \\
n_{sq} &=&\frac{1}{4}\left( \sqrt{\frac{\Delta X^{2}}{\Delta Y^{2}}}+\sqrt{%
\frac{\Delta Y^{2}}{\Delta X^{2}}}-2\right) \,,  \label{conn}
\end{eqnarray}
representing the average number of thermal and squeezed photons,
respectively. They measure the deviation of the actual state from the
minimum uncertainty one and its effective squeezing. In particular, the mean
total photon number is given by: 
\begin{equation}
N_{tot}=n_{sq}+n_{th}+2n_{sq}n_{th}\,,  \label{Ntot}
\end{equation}
while the variance of the generic quadrature $X_{\phi }$ reads: 
\begin{equation}
\Delta X_{\phi }^{2}=\frac{(2n_{th}+1)}{4}\left( 1+2n_{sq}+2\sqrt{\left(
1+n_{sq}\right) n_{sq}}\cos 2\phi \right) \,.  \label{quadrature}
\end{equation}

\begin{figure}[h]
\setlength{\unitlength}{1mm}
\par
\begin{center}
\includegraphics[width=0.7\textwidth]{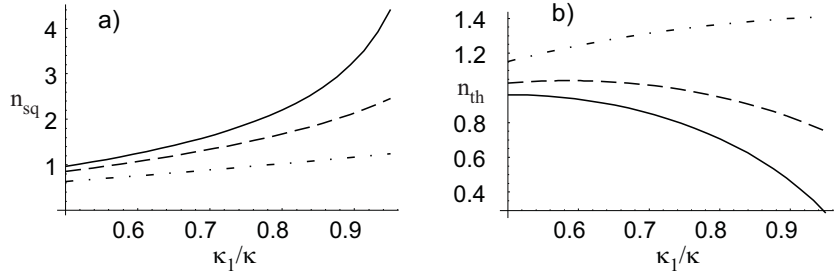}
\end{center}
\par
\vspace{-.5cm}
\caption{$n_{sq}$ (a) and $n_{th}$ (b) vs. coupling efficiency $\protect%
\kappa _{1}/\protect\kappa $ at half of the threshold power ($\mathcal{E}=0.5
$) for $\protect\omega =0$. The curves refer to three different detunings $%
\protect\psi =0.0,\,0.1,\,0.2$ (full, dashed and dot--dashed lines).}
\end{figure}

The photon numbers $n_{th}$ and $n_{sq}$ depend on the frequency offset $%
\omega$ and on the OPO parameters, namely, distance from the threshold ($%
\mathcal{E}$), escape efficiency ($\kappa _{1}/\kappa$), and cavity detuning
($\psi$). In Figs. 2--a and 2--b $n_{sq}$ and $n_{th}$ as functions of the
ratio $\kappa _{1}/\kappa$ and for three different detunings ($%
\psi=0.0,\,0.1,\,0.2$) have been plotted (OPO gain one-half of the threshold 
$\mathcal{E}=0.5$ and $\omega=0$). The detuning plays a more significant
role in proximity of $\kappa _{1}/\kappa \simeq 1$, that is for a
single-ended cavity configuration.

Since these states have been obtained by neglecting the time--dependent part
of both gain and detuning, they share the Gaussian statistics of the driving
fields $a^{in}_{1,2}$. They can be imagined as obtained by squeezing a
thermal state $\nu =(n_{th}+1)^{-1}\left[ n_{th}/(n_{th}+1)\right] ^{a^{\dag
}a}$, with $n_{th}$ the number of thermal photons (see Eq.(\ref{conn}--a)),
whose actual temperature is not necessarily coincident with the local one.
As a consequence the density matrix is: 
\begin{equation}
\varrho =S(\zeta )\nu S^{\dag }(\zeta )\,,  \label{rhoSTV}
\end{equation}
where $S(\zeta )=\exp \{\frac{1}{2}\zeta a^{\dag 2}-\frac{1}{2}\zeta ^{\ast
}a^{2}\}$ is the squeezing operator ($n_{sq}=\sinh ^{2}\left|\zeta\right|$ ) 
\cite{Marian,Man'ko}.

These STV states are described by a Gaussian Wigner function centered at the
origin: 
\begin{eqnarray*}
W\left( \alpha \right) &=&\frac{1}{2\pi \sqrt{\Delta X^{2}\Delta Y^{2}}}\exp
\left( -\frac{\Re\left[ \alpha \right] ^{2}}{2\Delta X^{2}}-\frac{\Im\left[
\alpha \right] ^{2}}{2\Delta Y^{2}}\right) \\
&=&\frac{2}{\pi }\int P\left( \beta \right) \exp \left( -2\left| \alpha
-\beta \right| ^{2}\right) d^{2}\beta \,,
\end{eqnarray*}
with $P\left( \beta \right) $ the corresponding P-representation: 
\[
P\left( \beta \right) =\frac{1}{2\pi \sqrt{\left( \Delta X^{2}-\frac{1}{4}%
\right) \left( \Delta Y^{2}-\frac{1}{4}\right) }}\exp \left( -\frac{\Re\left[
\beta \right] ^{2}}{2\left( \Delta X^{2}-\frac{1}{4}\right) }-\frac{\Im\left[
\beta \right] ^{2}}{2\left( \Delta Y^{2}-\frac{1}{4}\right) }\right) \,. 
\]
After propagation through a medium of transmittivity $T$ the density matrix 
\[
\varrho =\int P\left( \alpha \right) \left| \alpha \right\rangle
\left\langle \alpha \right| d^{2}\alpha 
\]
modifies as 
\[
\varrho _{T}=\int P\left( \alpha \right) \left| \sqrt{T}\alpha \right\rangle
\left\langle \sqrt{T}\alpha \right| d^{2}\alpha =\int P_{T}\left( \alpha
\right) \left| \alpha \right\rangle \left\langle \alpha \right| d^{2}\alpha
\,, 
\]
with 
\begin{eqnarray*}
\hspace{-1.5cm} P_{T}\left( \alpha \right) &=&\frac{1}{T}P\left( \frac{%
\alpha }{\sqrt{T}}\right) \\
\hspace{-1.5cm} &=&\frac{1}{2\pi \sqrt{\left( \Delta X_{T}^{2}-\frac{1}{4}%
\right) \left( \Delta Y_{T}^{2}-\frac{1}{4}\right) }}\exp \left( -\frac{\Re%
\left[ \alpha \right] ^{2}}{2\left( \Delta X_{T}^{2}-\frac{1}{4}\right) }-%
\frac{\Im\left[ \alpha \right] ^{2}}{2\left( \Delta Y_{T}^{2}-\frac{1}{4}%
\right) }\right) \,,
\end{eqnarray*}
and 
\[
\Delta X_{T}^{2}-\frac{1}{4}=T\left( \Delta X^{2}-\frac{1}{4}\right) \,. 
\]
$\Delta X^{2}-\frac{1}{4}$ is the deviation of the actual STV variance from
the vacuum state case (shot--noise). A similar expression is found for $%
\Delta Y_{T}^{2}$.

In principle, in the absence of multiple reflections within the sample, the
transmittivity $T$ is given by $T=T_{1}T_{slab}T_{2}$, where $T_{1}$ and $%
T_{2}$ are the Fresnel transmission coefficients at the input and output
faces of the sample respectively and $T_{slab}$ is the sample internal
transmittivity.

Next, introducing the subfixes $0$ and $T$ for labelling up-- and
down--stream quantities, respectively, for a generic quadrature $X_{\phi }$
the variance transforms as: 
\begin{equation}
\Delta X_{\phi ,T}^{2}-\frac{1}{4}=T\left( \Delta X_{\phi ,0}^{2}-\frac{1}{4}%
\right) \,.
\end{equation}
Accordingly, $T$ can be obtained by measuring the up-- and down--stream
quadrature variances: 
\begin{equation}
T=\frac{\Delta X_{\phi ,T}^{2}-\frac{1}{4}}{\Delta X_{\phi ,0}^{2}-\frac{1}{4%
}}\,.  \label{Tversus X}
\end{equation}
This relation suggests a simple way to measure $T$ through the deviations of
a generic quadrature from the vacuum noise level.

By means of Eqs.(\ref{quadrature}) and (\ref{Tversus X}), $T$ can be also
expressed as: 
\begin{equation}
T=\frac{(2n_{th,T}+1)\left( 1+2n_{sq,T}+2\sqrt{\left( 1+n_{sq,T}\right)
n_{sq,T}}\cos 2\phi \right) -1}{(2n_{th,0}+1)\left( 1+2n_{sq,0}+2\sqrt{%
\left( 1+n_{sq,0}\right) n_{sq,0}}\cos 2\phi \right) -1} \,.
\label{T versus n}
\end{equation}

On the other hand, $N_{tot}$ transforms proportionally to $T$ as for a
classical field: 
\begin{equation}
N_{tot,T}=T\,N_{tot,0}\,.  \label{newp3}
\end{equation}
Using Eq.(\ref{Ntot}) in the above expression and combining it with Eq.(\ref
{T versus n}) $n_{th,T}$ and $n_{sq,T}$ can be expressed in terms of $T$ and
of the initial values $n_{th,0}$ and $n_{sq,0}$: 
\begin{eqnarray}
\hspace{-2.5cm}2n_{th,T}+1 &=&\sqrt{\left[ 1-T+T(1+2n_{th,0})\left(
1+2n_{sq,0}\right) \right] ^{2}-\left[ 2T(1+2n_{th,0})\sqrt{\left(
1+n_{sq,0}\right) n_{sq,0}}\right] ^{2}}  \nonumber \\
\hspace{-2.5cm}2n_{sq,T}+1 &=&\frac{1-T+T(2n_{th,0}+1)\left(
1+2n_{sq,0}\right) }{2n_{th,T}+1}\,.  \label{nt&rt}
\end{eqnarray}
For the STV state used in the test discussed in Section 4 ($n_{th,0}=0.55$
and $n_{sq,0}=0.11$) $n_{th,T}$ and $n_{sq,T}$ are practically linear in $T$%
, that is: 
\begin{eqnarray}
\frac{n_{th,T}}{n_{th,0}} &=&A_{th}+B_{th}T  \nonumber \\
\frac{n_{sq,T}}{n_{sq,0}} &=&A_{sq}+B_{sq}T\,.  \label{linearnt&rt}
\end{eqnarray}
In Table \ref{expdata} the coefficients calculated by linearizing Eqs.(\ref
{nt&rt}) have been reported (first two columns) together with those obtained
experimentally (last two columns). For the sake of completeness the measured
ratio $N_{tot,T}/N_{tot,0}$ has been reported as well, in order to
evidenziate the agreement with the theoretical value of Eq.(\ref{newp3}).

\begin{table}[tbp]
\centering 
\begin{tabular}{|c|c|c|c|c|}
\hline
& $A$ & $B$ & $A^{(QHT)}$ & $B^{(QHT)}$ \\ \hline
$N_{tot}$ & $0$ & $1$ & $-0,05\pm 0,07$ & $1,1\pm 0,1$ \\ \hline
$n_{th}$ & $0.12$ & $0.89$ & $0.07\pm 0.05$ & $0.85\pm 0.07$ \\ \hline
$n_{sq}$ & $-0.12$ & $1.14$ & $-0.16\pm 0.05$ & $1.14\pm 0.07$ \\ \hline
\end{tabular}
\caption{Coefficients $A$ and $B$ computed by Eqs.(\ref{newp3},\ref
{linearnt&rt}) (left) and experimental ones measured by QHT (right). }
\label{expdata}
\end{table}

In the measurements discussed below, $T$ was determined through a direct
measurement of the parameters $n_{th}$ and $n_{sq}$ by QHT technique based
on pattern functions.

In conclusion, it is worth remarking that the above expressions of $T$ are
valid for Gaussian field quadratures. A Gaussian statistics follows from the
assumption of time independent gain and detuning of the OPO. In Section 4
the correctness of this assumption will be discussed for the used OPO by
measuring the deviations from the Gaussian statistics by means of the
kurtosis parameter $K_{\phi}$, vanishing for the Gaussian case, defined as: 
\begin{equation}
K_{\phi }=\frac{\overline{\Delta X_{\phi }^{4}}}{\left(\Delta X_{\phi
}^{2}\right)^{2}}-3 \,,  \label{kurtosis}
\end{equation}
$\overline{\Delta X_{\phi}^{4}}$ being the fourth order moment of $X_{\phi}$.

\section{Accuracy}

The limit of the uncertainty on the estimate of $T$ expressed by Eq.(\ref
{Tversus X}) depends on the confidence interval $\delta \left[ \Delta
X_{\phi }^{2}\right] $ in the measurement of $\Delta X_{\phi }^{2}$: 
\begin{equation}
\frac{\delta T}{T}=\frac{1}{\left\vert \Delta X_{\phi ,0}^{2}-\frac{1}{4}%
\right\vert }\sqrt{\frac{\delta \left[ \Delta X_{\phi ,T}^{2}\right] ^{2}}{%
T^{2}}+\delta \left[ \Delta X_{\phi ,0}^{2}\right] ^{2}}\,.
\label{accuracygen}
\end{equation}
Since 
\[
\delta \left[ \Delta X_{\phi }^{2}\right] =\sqrt{\frac{2}{N}}\Delta X_{\phi
}^{2}\,, 
\]
with $N$ the number of acquired data, the relative error on $T$ is given by: 
\begin{equation}
\frac{\delta T}{T}=\sqrt{\frac{2}{N}}\frac{1}{\left\vert \Delta X_{\phi
,0}^{2}-\frac{1}{4}\right\vert }\sqrt{\frac{1}{16}\left( 1-\frac{1}{T}%
\right) ^{2}+\frac{1}{2}\left\vert \Delta X_{\phi ,0}^{2}-\frac{1}{4}%
\right\vert \left( \frac{1}{T}+3+4\left\vert \Delta X_{\phi ,0}^{2}-\frac{1}{%
4}\right\vert \right) }\,.  \label{accuracyvacuum}
\end{equation}
This expression gives, for a given $T$, the relative error as a function of $%
N$ and $\Delta X_{\phi ,0}^{2}$ which in turns depends on the OPO working
condition, namely, distance from the threshold ($\mathcal{E}$), escape
efficiency ($\kappa _{1}/\kappa $), and cavity detuning ($\psi $). On the
other hand, the total number of photons $N_{ph}$ hitting the sample during
the measurement is: 
\begin{equation}
N_{ph}=N_{tot}\,N\,\kappa \,\tau _{s}\,,  \label{dose}
\end{equation}
with $\tau _{s}^{-1}$ the sampling rate and $N_{tot}$ given by Eq.(\ref{Ntot}%
).

\begin{figure}[h]
\setlength{\unitlength}{1mm}
\par
\begin{center}
\includegraphics[width=0.7\textwidth]{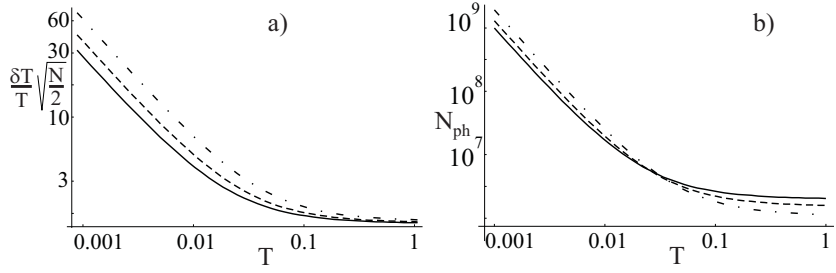}
\end{center}
\par
\vspace{-.5cm}
\caption{(a) Relative error $\frac{\protect\delta T}{T}\protect\sqrt{\frac{N%
}{2}}$ (Eq.(\ref{accuracyvacuum})) and (b) number $N_{ph}$ of photons
hitting the sample for $\frac{\protect\delta T}{T}=0.01$ (Eq.(\ref{dose})
and $\protect\kappa \protect\tau _{s}=6$) vs. transmittivity $T$. The plots
refer to $\protect\omega =\protect\psi =0$, at half the threshold ($\mathcal{%
E}=0.5$) and $\protect\kappa _{1}/\protect\kappa =0.5$, $0.75$, $1$)
(dot--dashed, dashed and full lines).}
\end{figure}

In Fig. 3--a $\frac{\delta T}{T}\sqrt{\frac{N}{2}}$ (see Eq. \ref
{accuracyvacuum})) has been plotted as a function of the transmittivity $T$
for $\omega =\psi =0$, $\mathcal{E}=0.5$, and three different escape
efficiencies ($\kappa _{1}/\kappa =0.5$, $\,0.75$, $1$). The relative error
increases for $T$ approaching zero. Fig.3--b gives the photon dose $N_{ph}$
(Eq. (\ref{dose})) necessary to obtain a relative error $\frac{\delta T}{T}%
=0.01$ for the parameters of Fig.3--a and $\kappa \tau _{s}=6$. The plot
evidentiates the increase of $N_{ph}$ by more than an order of magnitude for 
$T$ less than 0.01.

Instead of keeping $\phi $ constant during the acquisition of the $N$
samples, the angle can be varied uniformly in the interval $0\leq \phi \leq
2\pi $. The ensemble so obtained can be processed by means of QHT \cite{QHT}
for obtaining the field Wigner function. The tomographic processing can be
based on the so-called pattern function method, consisting in averaging the
pattern function $R_{\eta }[\hat{O}]\left( X_{\theta _{j}},\theta
_{j}\right) $ relative to an assigned operator $\hat{O},$ and having for
argument the $j$--th realization $X_{\theta _{j}}$ of $\hat{X}_{\theta }$
for the LO phase $\theta _{j}$, 
\begin{equation}
\langle \hat{O}\rangle =\frac{1}{N}\sum_{j=1}^{N}R_{\eta }[\hat{O}]\left(
X_{\theta _{j}},\theta _{j}\right) =\overline{R_{\eta }[\hat{O}]}\,.
\label{omean}
\end{equation}
The subfix ''$\eta $'' indicates the dependence of the pattern function on
the homodyne efficiency $\eta$.

For the operator $\hat{O}=\Delta X_{\phi }^{2}$ the confidence interval
provided by this method reads 
\begin{equation}
\delta _{QHT}\left[ \Delta X_{\phi }^{2}\right] =\frac{1}{\sqrt{N}}\sqrt{%
\overline{\Delta R^{2}\left[ \Delta X_{\phi }^{2}\right] }}\,,
\label{QHT error}
\end{equation}
with $\overline{\Delta R^{2}[\hat{O}]}=\overline{R_{\eta }^{2}[\hat{O}]}-%
\overline{R_{\eta }[\hat{O}]}^{2}$. Consequently Eq. (\ref{accuracygen}) is
still valid with $\delta \left[ \Delta X_{\phi }^{2}\right] $ replaced by $%
\delta _{QHT}\left[ \Delta X_{\phi }^{2}\right] $.

Next, taking into account the analytic expressions of $R_{\eta }^{2}[\Delta
X_{\phi }^{2}]$ and $R_{\eta }[\Delta X_{\phi }^{2}]$ \cite{rev}, it can be
shown that: 
\begin{equation}
\overline{\Delta R^{2}\left[ \Delta \hat{X}_{\phi }^{2}\right] }%
=C_{0}+C_{1}\cos \left( 2\phi \right) +C_{2}\cos \left( 4\phi \right) \,,
\label{variance R quadro}
\end{equation}
with the coefficients $C_{0}$, $C_{1}$ and $C_{2}$ given in Appendix. For
the variances $\Delta X^{2},\Delta Y^{2}$ relative to OPO devices similar to
that used in the experimental test, $\delta _{QHT}\left[ \Delta X_{\phi }^{2}%
\right] $ differs from $\delta \left[ \Delta X_{\phi }^{2}\right]$ only by
some percents. This means that collecting $N$ samples in the interval $(0,\,
2\pi)$ reduces the accuracy with respect to the constant phase case by only
a few percent.

Conventional measurements of $T$ using coherent CW probe beams and the
radiation power, $P$, as observable, are in some way corrupted by the
detector noise equivalent power ($NEP$), and the measurement error reads: 
\begin{equation}
\delta P=\sqrt{\hbar \omega_{0} B\;P}+NEP \,,  \label{error coh}
\end{equation}
with $\omega_{0} $ the radiation frequency, and $B$ the detection bandwidth.

Measuring $T$ as the ratio $P_{T}/P_{0}$ of the power down-- and up--stream
the sample the relative error is: 
\begin{equation}
\frac{\delta T}{T}=\frac{1}{SNR}\sqrt{\frac{1}{T^{2}}\left( 1+\sqrt{\frac{%
\hbar \omega _{0}B}{NEP}\,\frac{SNR\;T}{N}}\right) ^{2}+\left( 1+\sqrt{\frac{%
\hbar \omega _{0}B}{NEP}\,\frac{SNR}{N}}\right) ^{2}}\,,  \label{IconDark}
\end{equation}
with $SNR=P_{0}/NEP$ and $N$ the number of data.

The total number of photons (see Eq. (\ref{dose})) passing through the
sample during the measurement interval is now given by 
\begin{equation}
N_{ph}=SNR\frac{NEP}{\hbar \omega _{0}}N\tau _{s}  \label{dosebis}
\end{equation}
so that, the factor $\frac{\hbar \omega _{0}B}{NEP\,N}$ in Eq. (\ref
{IconDark}) can be replaced by $\frac{SNR\,B\tau _{s}}{N_{ph}}$ (with $B\tau
_{s}>1$). Then the ratio $\frac{B\tau _{s}}{N_{ph}}$ is a function of $\frac{%
\delta T}{T}$, $T$ and $SNR$. Using for $SNR$ the limiting value 
\[
SNR\geq \frac{T}{\delta T}\sqrt{\frac{1}{T^{2}}+1}\,, 
\]
the plot of Fig. 4 representing $N_{ph}$ vs. $T$ for $\frac{\delta T}{T}=0.01
$ and $B\tau_{s}=10$ has been obtained. Comparing it with Fig. 3--b it
appears evident the much lower photon dose required by the present method.

\begin{figure}[h]
\setlength{\unitlength}{1mm}
\par
\begin{center}
\includegraphics[width=0.5\textwidth]{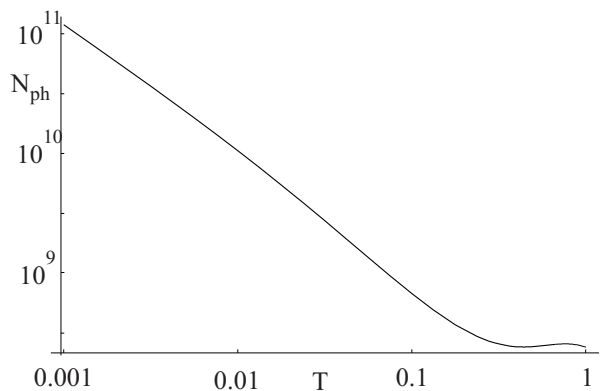}
\end{center}
\par
\vspace{-.5cm}
\caption{Photon dose passing through the sample ($N_{ph}$) vs. $T$ in case
of standard transmission measurements. The curve refer to $B\protect\tau%
_{s}=10$ and $\frac{\protect\delta T}{T}=0.01$.}
\end{figure}

In case a coherent beam ($a_{\alpha }$) is mixed with a squeezed thermal
vacuum one ($a_{STV}$), as in Ref. \cite{polzik}, the total field is
described by: 
\[
a_{tot}=e^{i\theta }a_{STV}+a_{\alpha }\,, 
\]
with $\theta $ their locked phase difference.

It can be shown that the addition of the squeezed component modifies
slightly Eq. (\ref{error coh}) with $B$ replaced by 
\[
B_{eff}=B\,\left( 1+n_{sq}+n_{th}+2n_{sq}n_{th}\,+\sqrt{\left(
1+n_{sq}\right) n_{sq}}\cos 2\theta \right) \,.
\]
For $\cos 2\theta =-1$ and $\sqrt{\left( 1+n_{sq}\right) n_{sq}}%
>n_{sq}+n_{th}+2n_{sq}n_{th}$ the squeezed vacuum component reduces the
effective detector bandwidth. The reduction of $B$ (typically $%
B_{eff}\gtrsim .5B$) implies a proportional decrease of $N_{ph}$ for
assigned $\frac{\delta T}{T}$ and $T$.

\section{The experiment}

The reliability and accuracy of the method were tested with a sample of
variable transmittivity. The $T$ values obtained via QHT were compared to
those measured, with an accuracy of $10^{-4}$, with standard techniques
employing 1 mW coherent beam at $\lambda =1064$ nm. A schematic of the
experimental set--up is shown in Fig. 1.

STV states were generated by a degenerate type--I OPO and characterized by a
homodyne detector, both described in details in Ref. \cite{opticsexpress}.
In the present case, cavity mirrors were adjusted in such a way as to have a
cavity linewidth of 15 MHz.

The OPO output was propagated through a variable neutral density filter,
which changes $T$ without introducing misalignment, and keeping homodyne
visibility at a constant value. The transmittivity $T$ was varied between $%
0.45$ and $1$ in discrete steps. The beam passing through the non--absorbing
zone ($T=1$) of the filter was used as a reference state. The field leaving
the absorber was sent to an homodyne detector with an overall efficiency of $%
\eta =0.88\pm 0.02$. The average electrical signal level at the homodyne
output was 15 dB higher than the electronic noise.

Tomographic data were acquired by sampling the homodyne signal. To avoid any
effect of the laser technical noise on the measurement, data sampling was
performed by mixing the homodyne current with a sinusoidal signal of
frequency $\Omega =5$ MHz. Then, the resulting current was low--pass
filtered, with a cut--off frequency of 2.5 MHz, and 10$^{6}$ samples were
collected with at 2.5 Msample/s ($\tau _{s}=400$ ns) in order to pick-up
statistically independent data.

\begin{figure}[h]
\setlength{\unitlength}{1mm}
\par
\begin{center}
\includegraphics[width=0.7\textwidth]{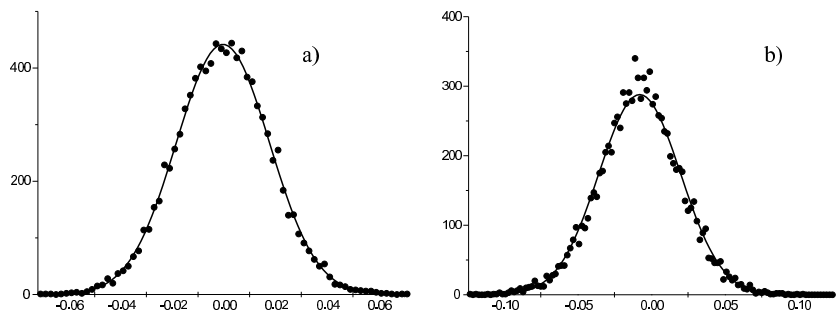}
\end{center}
\par
\vspace{-.5cm}
\caption{Distribution of $X$ values measured for $\mathcal{E}=0.5$ a) and
0.95 b). The kurtosis $K_{0}$ (Eq. (\ref{kurtosis})) is respectively equal
to 0.005 and 0.5. Full lines represent Gaussian with the same mean and
variance.}
\end{figure}

Fixing $\mathcal{E}=0.50$ the reference STV state had $N_{tot,0}=0.79\pm 0.06
$, $n_{th,0}=0.55\pm 0.02$ and $n_{sq,0}=0.11\pm 0.01$, corresponding to a
photon flux of $10^{7}$ s$^{-1}$. For this state it resulted $K_{\phi
}\lesssim 0.01$ (see Fig. 5--a and Eq.(\ref{kurtosis})) for any $\phi $,
thus indicating that the corresponding quadrature statistics was very close
to the Gaussian one.

In order to reduce the influence of residual fluctuations of the STV state,
each experimental point was averaged over multiple ($\sim $5) tomographic
acquisitions. In the present conditions the QHT error was negligible with
respect to the standard deviations of the STV state parameters.

To assess the robustness of the method, the transmittivity, $T_{QHT}$,
obtained by tomographic reconstruction was compared with the corresponding
value, $T_{st}$, provided by standard intensity measurements.

\begin{figure}[h]
\setlength{\unitlength}{1mm}
\par
\begin{center}
\includegraphics[width=0.5\textwidth]{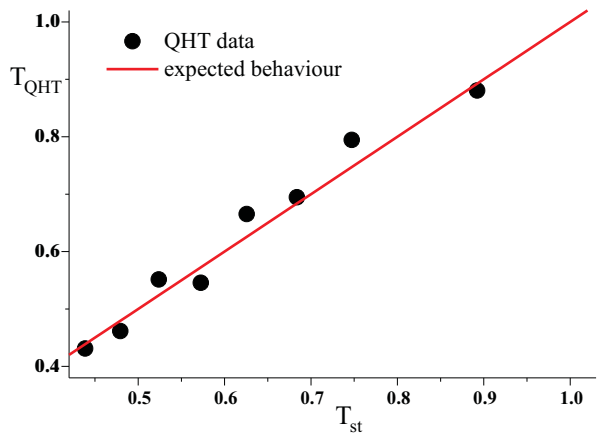}
\end{center}
\par
\vspace{-.5cm}
\caption{$T_{QHT}$ vs. $T_{st}$. Experimental points are plotted together
with the expected behaviour of Eq. (\ref{newp3}) (straight line).}
\end{figure}

In Fig. 6, $T_{QHT}=N_{tot,T}/N_{tot,0}$ (see Eq.(\ref{newp3})) was plotted
vs. $T_{st}$ together with the expected behavior $T_{QHT}=T_{st}$ (straight
line). A linear regression of the data with $%
T_{QHT}=A_{tot}^{(QHT)}+B_{tot}^{(QHT)}T_{st}$, gave $A_{tot}^{(QHT)}=-0.05%
\pm 0.07$ and $B_{tot}^{(QHT)}=1.1\pm 0.1$ in good agreement with the
expected values of $A_{tot}=0$ and $B_{tot}=1$ respectively.

In order to estimate $T_{QHT}$ through other quantities, the measured value
of $n_{sq,T}/n_{sq,0}$ versus $T_{st}$ was plotted in Fig. 7 together with
the linear approximation of Eq. (\ref{linearnt&rt}--b). Linear regression on
experimental data gave $A_{sq}^{(QHT)}=-0.16\pm 0.05$ and $%
B_{sq}^{(QHT)}=1.14\pm 0.07$, values in good agreement with $A_{sq}=-0.12$, $%
B_{sq}=1.14$.

Each experimental point of Fig. 7 represents an average value obtained over
multiple acquisitions. In the inset the different values of $T_{QHT}$,
corresponding to four acquisitions at $T_{st}=0.64$ are reported. The bar
indicates the quantum limit error, calculated by using Eq. (\ref
{accuracyvacuum}). As it can be seen, all the points are spread over a range
comparable to the quantum limit.

\begin{figure}[h]
\setlength{\unitlength}{1mm}
\par
\begin{center}
\includegraphics[width=0.5\textwidth]{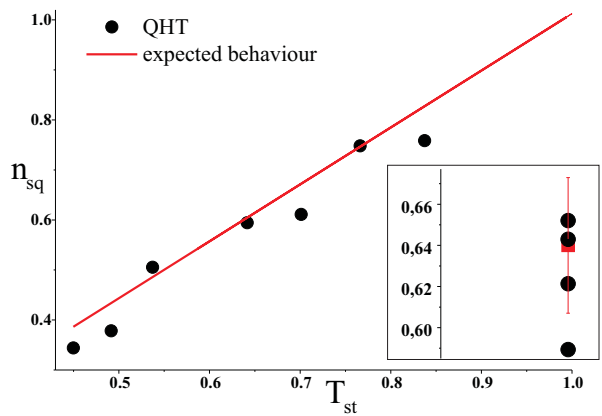}
\end{center}
\par
\vspace{-.5cm}
\caption{$n_{sq,T}$ vs. $T_{st}$. Experimental points are compared with Eq.(%
\ref{linearnt&rt}b). The points in the inset are relative to four
measurements for an attenuator transmittivity equal to 0.64 while the error
bar has been calculated by using Eq. (\ref{accuracyvacuum}).}
\end{figure}

In Fig. 8 the shadowed area represents the quantum limit for the accuracy
vs. $T$ (see Eq.(\ref{accuracyvacuum})) for the present experimental
conditions and $N=10^{4}$. The accuracy width is almost constant in the
tested range of $T$ while it deteriorates for low transmittivity, as
expected.

\begin{figure}[h]
\setlength{\unitlength}{1mm}
\par
\begin{center}
\includegraphics[width=0.5\textwidth]{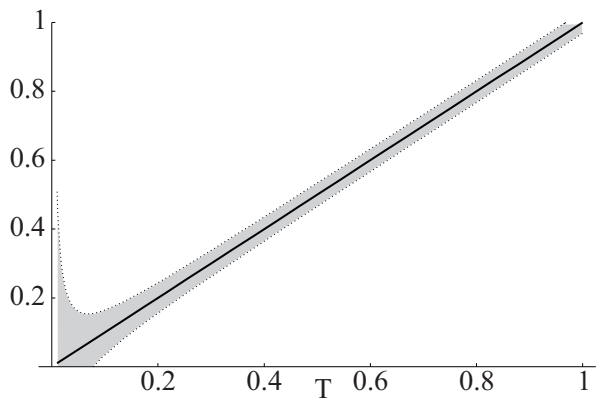}
\end{center}
\par
\vspace{-.5cm}
\caption{Quantum limit (shadowed area) of the relative error on $T$ for the
experimental case discussed in the text and $N=10^{4}$.}
\end{figure}

Finally, an identical behavior was observed for $n_{th,T}/n_{th,0}$ (not
plotted) resulting in $A_{th}^{(QHT)}=0.07\pm 0.05$, $B_{th}^{(QHT)}=0.85\pm
0.07$ ($A_{th}=0.12$, $B_{th}=0.89$).

A summary of the experimental findings is reported in Table \ref{expdata}.

The photon flux at the OPO output $F=N_{tot}/\tau $, with $\tau $ the cavity
photon lifetime, was less than $10^{7}$ s$^{-1}$, for $N_{tot}\lesssim 0.7$
and $\tau\approx 6.6\times 10^{-8}$, corresponding to an optical power $%
\lesssim 4.2$ pW. The method was tested for different input states, by
varying $\mathcal{E}$ and hence the photon flux by showing a good
reliability down to a photon flux $F\sim 5\times 10^{6}$ s$^{-1}$ (i.e. $%
\sim $2.2 pW and $N_{tot,0}=0.37$).

With $N=10^{6}$ $\delta_{QHT}\left[\Delta X^{2}\right]\sim 1.\,3\times
10^{-3}$ and $\delta _{QHT}\left[ \Delta Y^{2}\right]\sim 0.8\times 10^{-3}$
corresponding to $\delta T/T$ $\sim 0.0024$ and $\sim 0.056$ for $T=1$.
These QHT estimates were slightly less accurate than those one could obtain
by concentrating $N/2$ data on $X$ and $N/2$ on $Y$ quadratures and
computing their variances.

\section{Conclusions}

A scheme for measuring the optical transmittivity of a sample by using
squeezed vacuum radiation has been illustrated. Main advantage of this
method is a number of photons hitting the sample during the measurement some
orders of magnitude smaller than that relative to standard techniques based
on intensity measurements of coherent beams.

The core of the method consists in the measurement of the variance $\Delta
X_{\phi }^{2}$ of a generic quadrature of a squeezed vacuum field, generated
by a below threshold OPO and passing through the sample under investigation.
The quadrature is measured by a homodyne detector. In the simplest
implementation $\Delta X_{\phi }^{2}$ is obtained by averaging the squared
samples $X_{\phi }$ relative to a constant phase $\phi$. In the test
described in the paper $X_{\phi}$ has been obtained by scanning the interval 
$\phi \in \left(0,\,2\pi \right)$. This approach has been preferred since it
provides a complete reconstruction of the squeezed vacuum Wigner function.

Essential to this scheme is the assumption of Gaussian statistics for the
squeezed vacuum field. This property has been checked on the recorded
samples relative to a given phase and confirmed by the field Wigner function.

The accuracy of this method has been compared with that based on absorption
of coherent beams (with and without a squeezed vacuum component) as a
function of sample transmittivity, number of data and detection bandwidth.
In the case the number of photons interacting with the sample during the
measurement is an important parameter, the proposed method is the most
accurate.

The experimental tests have shown that, for photon fluxes of the order of
few pW (at 1064 nm), the accuracy is of the order of the quantum limit, that
is the method does not suffer substantially from other technical noise
sources

\section*{Acknowledgments}
This work has been supported by MIUR through the project PRIN-2005024254.

\section*{Appendix}

The coefficients $C_{0}$, $C_{1}$ and $C_{2}$ of Eq. ({\ref{variance R
quadro}}) are given by: 
\begin{eqnarray*}
\hspace{-2.5cm} C_{0} &=&\frac{1}{4}\left[ \frac{27}{2}\left( \Delta
X^{4}+\Delta Y^{4}\right) +9\Delta X^{2}\Delta Y^{2}+\left( 1-\frac{3}{\eta }%
\right) \left( \Delta X^{2}+\Delta Y^{2}\right) +\frac{1}{4}\left( \frac{3}{%
\eta ^{2}}-\frac{2}{\eta }+1\right) \right] \\
\hspace{-2.5cm} C_{1} &=&\frac{1}{2}\left( \Delta X^{2}-\Delta Y^{2}\right) %
\left[ 3\left( \Delta X^{2}+\Delta Y^{2}\right) -1\right] \\
\hspace{-2.5cm} C_{2} &=&\frac{3}{8}\left( \Delta X^{2}-\Delta Y^{2}\right)
^{2}\,.
\end{eqnarray*}

\end{document}